# Spherical Harmonic Analysis of the Angular Distribution of GRBs

Max Tegmark[1,2], Dieter H. Hartmann[3], Michael S. Briggs[4], and Charles. A. Meegan[5]

[1] *Max-Planck-Institut für Physik, Föhringer Ring 6,D-80805 München*
[2] *Max-Planck-Institut für Astrophysik, D-85740 Garching*
[3] *Dept. of Physics and Astronomy, Clemson University, Clemson, SC 29634*
[4] *Dept. of Physics, University of Alabama, Huntsville, AL 35899*
[5] *NASA/Marshall Space Flight Center, Huntsville, AL 35812*

We compute the angular power spectrum $C_\ell$ of the BATSE 3B catalog, and find no evidence for clustering on any scale. These constraints bridge the entire range from small scales, probing source clustering and repetition, to large scales constraining possible Galactic anisotropies, or those from nearby cosmological large scale structures.

## I. INTRODUCTION

The observed angular distribution of $\gamma$-ray bursts (GRBs) is isotropic, while their brightness distribution shows a reduced number of faint events. These observations favor a cosmological burst origin. Clustering of bursts could be evidence of actual clustering of sources or of repeated emission. Repetition would call into question the viability of many cosmological burst models. Anisotropies manifest themselves on different angular scales and with different magnitudes. Galactic features cause large-scale distortions, while true repetition would affect small scales. For large-scale signatures, we search for excesses of sources towards some direction or a concentration towards some plane in the sky, i.e., we seek a dipole- or quadrupole moment. It is now common practice to apply both coordinate-free and galactic tests (1). Dipole- and quadrupole measures were sufficient when sample sizes were small. Now an extension of moment methods to higher orders is needed. Low order multipoles are not sensitive to instrumental smearing, but higher harmonics are. If associated with galaxies, we expect clustering on very small scales. If bursts repeat, we expect clustering at $\theta=0$. Both effects are diluted by localization uncertainties, and angular power is transferred from small (or zero) angular scales to a scale given by the detector response. One tool for the analysis of source clustering is the two-point correlation function (2), which is related to the power spectrum through a Fourier transform (4).

## II. METHOD

We model the GRB distribution as a 2D stochastic point process $n(\hat{\mathbf{r}}) = \sum_i \delta(\hat{\mathbf{r}}, \hat{\mathbf{r}}_i)$ with intensity (average point density per steradian) $\lambda(\hat{\mathbf{r}})$. Here $\delta$ denotes the 2D Dirac delta function, and the unit vectors $\hat{\mathbf{r}}_i$ correspond to the various GRB positions. If we had detected a nearly infinite number of bursts, then the function $\lambda(\hat{\mathbf{r}})$ would be known with great accuracy, and the only source of errors when computing its power spectrum would be cosmic variance. Since in practice we have only a finite number of bursts (1122 for 3B), our estimates of $\lambda$ include shot noise. A Poisson process satisfies the expectation value equations

$$\langle n(\hat{\mathbf{r}}) \rangle = \lambda(\hat{\mathbf{r}}), \tag{1}$$

and

$$\langle n(\hat{\mathbf{r}}) n(\hat{\mathbf{r}}') \rangle = \lambda(\hat{\mathbf{r}}) \lambda(\hat{\mathbf{r}}') + \delta(\hat{\mathbf{r}}, \hat{\mathbf{r}}') \lambda(\hat{\mathbf{r}}). \tag{2}$$

Here $\lambda$ is itself a random field, $\lambda(\hat{\mathbf{r}}) = \bar{n}(\hat{\mathbf{r}})[1 + \Delta(\hat{\mathbf{r}})]$, where the underlying density fluctuations $\Delta$ are modeled as a Gaussian random field. The function $\bar{n}$, which we will refer to as the *exposure function*, is the number of bursts per steradian expected a priori, not the number density actually observed. In other words, $\bar{n}(\hat{\mathbf{r}})$ is proportional to the exposure time in the sky direction $\hat{\mathbf{r}}$. We assume that $\langle \Delta(\hat{\mathbf{r}}) \rangle = 0$ and that the statistical properties of the field $\Delta$ are isotropic, which means that if we expand it in spherical harmonics as

$$\Delta(\hat{\mathbf{r}}) = \sum_{\ell=0}^{\infty} \sum_{m=-\ell}^{\ell} a_{\ell m} Y_{\ell m}(\hat{\mathbf{r}}), \tag{3}$$

then

$$\langle a_{\ell m} a_{\ell' m'} \rangle = \delta_{\ell \ell'} \delta_{m m'} C_\ell, \tag{4}$$

where the coefficients $C_\ell$ are known as the *angular power spectrum*. There are thus two separate random steps involved in generating $n$: first the generation of the smooth field $\Delta$, then the Poissonian distribution of points.

Given the field $n(\hat{\mathbf{r}})$, we wish to estimate the coefficients $a_{\ell m}$. We define them as

$$\tilde{a}_{\ell m} \equiv \int Y_{\ell m}(\hat{\mathbf{r}}) \frac{n(\hat{\mathbf{r}})}{\bar{n}(\hat{\mathbf{r}})} d\Omega - \delta_{\ell 0} \delta_{m 0} \sqrt{4\pi}. \tag{5}$$

We now compute the statistical properties of these estimates. By substitution we obtain

$$\langle \tilde{a}_{\ell m} \rangle = \int Y_{\ell m}(\hat{\mathbf{r}}) d\Omega - \delta_{\ell 0} \delta_{m 0} \sqrt{4\pi} = 0, \tag{6}$$

*i.e.*, the expectation values vanish. Since the expectation values of the true coefficients $a_{\ell m}$ vanish as well, this means that our estimates are unbiased.

Using the expressions above, we find that the correlation between two multipole estimates is

$$\langle \tilde{a}_{\ell m} \tilde{a}_{\ell' m'} \rangle = \int \int Y_{\ell m}(\hat{\mathbf{r}}) Y_{\ell' m'}(\hat{\mathbf{r}}') \left[ \langle \Delta(\hat{\mathbf{r}}) \Delta(\hat{\mathbf{r}}') \rangle + \frac{1}{\bar{n}(\hat{\mathbf{r}})} \delta(\hat{\mathbf{r}}, \hat{\mathbf{r}}') \right] d\Omega d\Omega', \quad (7)$$

which reduces to

$$\langle \tilde{a}_{\ell m} \tilde{a}_{\ell' m'} \rangle = \delta_{\ell \ell'} \delta_{m m'} C_\ell + \int \frac{Y_{\ell m}(\hat{\mathbf{r}}) Y_{\ell' m'}(\hat{\mathbf{r}})}{\bar{n}(\hat{\mathbf{r}})} d\Omega. \quad (8)$$

Defining the quantities

$$\tilde{C}_{\ell m} \equiv \tilde{a}_{\ell m}^2 - b_{\ell m}, \quad (9)$$

we find that they are unbiased estimates if we choose the *bias correction* as

$$b_{\ell m} \equiv \int \frac{Y_{\ell m}^2(\hat{\mathbf{r}})}{\bar{n}(\hat{\mathbf{r}})} d\Omega. \quad (10)$$

If $\bar{n}$ is constant, then the bias correction becomes simply $b_{\ell m} = 1/\bar{n}$, independent of $\ell$ and $m$. The $\tilde{C}_{\ell m}$ are thus good estimates of $C_\ell$ for each $m$-value separately. To reduce error bars, we estimate power by averaging the $\tilde{C}_{\ell m}$:

$$\tilde{C}_\ell \equiv \frac{1}{2\ell + 1} \sum_{m=-\ell}^{\ell} \tilde{C}_{\ell m}. \quad (11)$$

Defining $b$ to be the average of the bias corrections $b_{\ell m}$, we find that $b$ is in fact independent of $\ell$, and obtain

$$b \equiv \frac{1}{2\ell + 1} \sum_{m=-\ell}^{\ell} b_{\ell m} = \frac{1}{4\pi} \int \frac{d\Omega}{\bar{n}(\hat{\mathbf{r}})}, \quad (12)$$

*i.e.*, $b$ is just the spherical average of $1/\bar{n}$.

It is straightforward to include the effects of position errors in the formalism, which is described in a more detailed ApJ version of this paper (5). We model the BATSE beam function as a Fisher function

$$B(\hat{\mathbf{r}} \cdot \hat{\mathbf{r}}') = \frac{\exp[\sigma^{-2} \hat{\mathbf{r}} \cdot \hat{\mathbf{r}}']}{4\pi \sigma^2 \sinh[\sigma^{-2}]}, \quad (13)$$

characterized by a location error $\sigma$. This is a spherical version of the Gaussian distribution, and reduces to

$$B(\cos\theta) \approx \frac{\exp\left[-\frac{1}{2}\frac{\theta^2}{\sigma^2}\right]}{2\pi \sigma^2} \quad (14)$$

when $\sigma \ll 1$ radian $\approx 60°$. The Fisher function has the advantage that it is correctly normalized (its integral over the sphere is unity) for arbitrarily large angles $\sigma$, which is not the case for the plane Gaussian. In addition to statistical position errors we include (in quadrature) a $1.6°$ systematic uncertainty. This value is significantly lower than the $4°$ of earlier catalogs, allowing us to extend spherical harmonic analysis to $\ell \sim 60$ before localization uncertainties wash out intrinsic angular power.

## III. RESULTS AND DISCUSSION

The power spectrum $\tilde{C}_\ell$ of the the 3B data (3) is shown in Figure 1. There is no evidence of deviations from isotropy on any angular scale. If the gamma-ray bursts are completely uncorrelated, the points should scatter symmetrically around zero, with about 68% in the shaded region. Since all power is by definition positive, the presence of any type of clustering would shift the distribution upwards, leading to a positive excess. A monopole $C_0/4\pi = 0.0001$ corresponds to a fluctuation of $\sqrt{0.0001} = 1\%$ in the average burst density. Likewise, $[C_\ell/4\pi]^{1/2}$ can be interpreted as the density fluctuation on the angular scale $\theta \approx 60°/\ell$. The size of the error bars (the height of the shaded region) is readily understood. For $\ell = 0$, all $N = 1122$ bursts carry equal weight, so apart from a factor of $\sqrt{2}$, the shot noise gives just the familiar Poisson variance $1/N$. As $\ell$ increases, the error bars become smaller since $(2\ell + 1)$ independent modes are being averaged. Since the weighting scheme loosely speaking only obtains information on $C_\ell$ from bursts better localized than $60°/\ell$, the $1/N$ shot noise finally causes the error bars to grow with $\ell$ again, since the effective number of bursts $N$ decreases.

Although the angular power spectrum $C_\ell$ provides a useful measure of the amount of clustering on different angular scales, it does not contain any information about the relative *phases* of the different multipoles $a_{\ell m}$. The loss of phase-information means that although the power spectrum may tell us that there is extra power on some scale, it does not tell us anything about where in the sky this power is coming from. Fortunately, this type of information is easy to extract through definition of a map $x_\ell(\hat{\mathbf{r}})$, the *multipole map* corresponding to multipole $\ell$, as the sky map

$$x_\ell(\hat{\mathbf{r}}) \equiv \sum_{m=-\ell}^{\ell} \tilde{a}_{\ell m} Y_{\ell m}(\hat{\mathbf{r}}), \qquad (15)$$

The multipole information that our SHA extracts from the data, as plotted in Figure 1, places sharp quantitative limits on repetition. Suppose that a fraction $f$ of all observed bursts can be labeled as repeaters that are observed to burst $\nu$ times each. Application of an SHA-based technique to test this two-parameter family of models against the BATSE 3B data shows that all models with $(\nu - 1)f \geq 0.05$ are ruled out at 99% confidence (6), as compared to the best previous 99% limit $(\nu - 1)f \geq 0.27$. Thus even a cluster of 6 events from a single source would have caused excess power above that present in the 3B catalog.

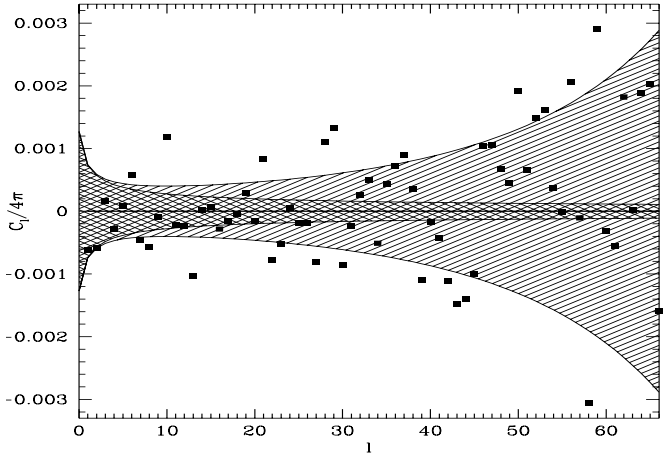

**FIG. 1.** The shot-noise corrected angular power spectrum of 3B (solid squares). The shaded region shows the $1\sigma$ error bars. Any type of clustering would drive the measured points upward. The double-shaded region shows what the errors would be without localization uncertainties.

In summary, multipole expansion of the projected distribution of GRBs does not show evidence for clustering on any angular scale. This argues against the recurrence of a substantial fraction of burst sources (6) and against any source population with strong intrinsic anisotropies. The remarkable degree of isotropy of GRBs severely constrains any burst model that invokes traditional geometric features of the Milky Way (disk, bulge, or halo).

This work was supported by NASA under grant NAG 5-1578 and by Deutsche Forschungsgemeinschaft under grant SFB-375.